\documentclass{emulateapj}
\shorttitle{Hard X-RAY EMISSION FROM CLASSICAL NOVAE}
\shortauthors{SUZUKI\&SHIGEYAMA}
\begin{document}
\title{COMPTON DEGRADATION OF GAMMA-RAY LINE EMISSION FROM RADIOACTIVE ISOTOPES IN THE CLASSICAL NOVA V2491 CYGNI}
\author{AKIHIRO SUZUKI\altaffilmark{1,2} and TOSHIKAZU SHIGEYAMA\altaffilmark{1}}
\altaffiltext{1}{Research Center for the Early Universe, School of Science, University of Tokyo, Bunkyo-ku, Tokyo 113-0033, Japan.}
\altaffiltext{2}{Department of Astronomy, School of Science, University of Tokyo, Bunkyo-ku, Tokyo 113-0033, Japan.}
\begin{abstract}
To account for the non-thermal emission from the classical nova V2491 Cygni, we perform a series of numerical calculations of radiative transfer of $\gamma$-ray photons from the radioactive isotope $^{22}$Na in the matter ejected from a white dwarf. 
Using a simple wind model for the dynamical evolution of the ejecta and a monte-carlo code, we calculate radiative transfer of the $\gamma$-ray photons in the ejecta. 
Repeated scattering of the $\gamma$-ray photons by electrons in the ejecta, i.e., Compton degradation, results in an extremely flat spectrum in the hard X-ray range, which successfully reproduces the observed spectrum of the X-ray emission from V2491 Cygni. 
The amount of the isotope $^{22}$Na synthesized in the ejecta is required to be $3\times 10^{-5}\ M_\odot$ to account for the flux of the hard X-ray emission.  
Our model indicates that the ejecta become transparent to the $\gamma$-ray photons within several tens days. 
Using the results, we briefly discuss the detectability of the $\gamma$-ray line emission by the {\it INTEGRAL} gamma-ray observatory and the {\it Fermi} Gamma-ray Space Telescope. 
\end{abstract}

\section{INTRODUCTION\label{intro}}
A classical nova originates from the thermonuclear runaway of hydrogen-rich gas having been accreted by a white dwarf in a close binary system. 
Due to the enormous amount of the released nuclear energy, the envelope of the white dwarf is blown up to the interstellar space. 
The expanding ejecta generate strong shock waves in the interstellar medium and the ejecta. 
The shocked matter radiates in optical to X-ray energy ranges in a similar manner to its more energetic counterpart, supernova explosions. 
From both observational and theoretical sides, the emission mechanism and the evolution of the ejecta have been investigated. 

The classical nova V2491 Cygni, which was discovered by \cite{n08} on 2008 April 10.728 UT, is one of the most outstanding examples among X-ray observed classical novae. 
The {\it Swift} satellite observed V2491 Cygni on the 5th day after the discovery and detected X-ray emission\citep{k08}. 
Then, the target-of-opportunity observations by the {\it Suzaku} observatory were performed on days 9 and 29 after the discovery. 
\cite{t09} reported an extremely hard emission in the spectrum taken on day 9. 
The spectrum is well fitted by a thermal emission from an optically thin plasma combined with a hard power-law component.  
The best-fit temperature and the photon index are $k_\mathrm{B}T=2.9^{+4.3}_{-2.6}$ keV and $\Gamma=0.1\pm 0.2$, respectively. 
On the other hand, the hard emission became absent on day 29. 
Since the bremsstrahlung model cannot explain the 
spectrum, they concluded that the 
emission is of non-thermal origin. 
However, the emission mechanism of the non-thermal photons remains unclear.

\cite{t08} acquired optical spectra of V2491 Cygni on days 1 and 3, and found P Cygni profiles of H$\beta$ and H$\gamma$ with the velocity of $\sim 4000$ km s$^{-1}$. 
From the similarity of the spectrum of V2491 Cygni to those of recurrent novae, such as U Sco \citep{m99}, \cite{t08b} argued that V2491 Cygni is a recurrent nova. 
They also reported that the light curve of V2491 Cygni showed very fast evolution, which indicates that the nova originates from a massive white dwarf. 
\cite{hk09} suggest that the white dwarf mass is $1.3\pm0.02\ M_\odot$ by using light-curve fittings based on the optically thick wind model \citep{kh94}. 
The massive white dwarf origin and the large velocity ($\sim 4000$ km s$^{-1}$) imply the outburst on the surface of an ONeMg white dwarf \citep[e.g.,][]{jh98}.
Interestingly, a variable X-ray source has been observed at the position of V2491 Cygni \citep{ik08}. 
Its hard spectrum was reminiscent of persistent emissions from magnetic cataclysmic variables \citep{i08}. 
However, the lack of periodically varying X-ray emissions from V2491 Cygni in the later phase contradicts the presence of strong magnetic fields \citep{p10}.

In this Letter, we show that the non-thermal emission can be explained by Compton degradation of $\gamma$-ray line emission from the radioactive isotope $^{22}$Na synthesized in V2491 Cygni. 
In the early phase of the evolution of the ejecta, the $\gamma$-ray photons are scattered by electrons in the optically thick ejecta and eventually degraded to X-ray photons. 
The emission of $\gamma$-ray photons from radioactive isotopes and its importance in the production of hard X-ray photons have been pointed out by many authors \citep{c74,c81,l92,g98}. 
Actually, from supernova 1987A, hard X-rays as a result of Compton degradation of $\gamma$-rays originated from $^{56}$Co were predicted based on theoretical models \citep{i87,x88,e88} and detected by Ginga \citep{m88a,m88b} and Kvant \citep{s87}. 
Here, we argue that a similar phenomenon due to another radioactive isotope $^{22}$Na occurs in this nova. 

The production of radioactive isotopes in the nova nucleosynthesis has been investigated by many authors\citep{s78,w90,n91,c95,p95,jh98,w99,s00}. 
However, the amount of the produced $^{22}$Na remains uncertain due to uncertainties in reaction rates used in the nuclear reaction network\citep{i02,j04,d04,c07,s10}. 
Although the extended $\gamma$-ray line emission, such as the $1.27$ MeV line from $^{22}$Na, from the Galactic bulge is detected and considered to be integrated emission from individual novae, no detection of $\gamma$-ray line emissions from individual novae has been reported so far\citep[e.g.,][]{l88,i95,i05}. 

We calculate spectra based on a simple wind model for the dynamical evolution of the ejecta and compare the resultant spectra with observations of V2491 Cygni. 
Radiative transfer of $\gamma$-ray photons in the ejecta is treated in the test-particle limit. 
We use $10.5$ kpc as the distance to V2491 Cygni, following the previous works \citep[see,][]{h08}.  
In Section 2, we describe our model for the ejecta and the procedure of the radiative transfer calculation. 
The resultant spectra are shown in Section 3. 
In Section 4, we discuss the detectability of the $\gamma$-ray line emission. 
Finally, Section 5 concludes this Letter.  

\section{FORMULATION\label{formulation}}
In this section, we describe a procedure to calculate a spectrum of degraded $\gamma$-ray line emission from $^{22}$Na. 
We numerically deal with the radiative transfer problem, because we need to treat photons with the energies higher than the electron rest energy and include effects of photoelectric absorption. 
We use a monte-carlo radiative transfer code having been developed by the present authors\citep{ss10}. 
We explain only modifications to the original code in subsections 2.2 and 2.3. 

\subsection{Ejecta model}
We assume that the envelope of the white dwarf is ejected at a constant mass-loss rate with a uniform velocity $v_\mathrm{ej}$ for a time interval $\tau$. 
The resultant electron number density $n_\mathrm{e}(r,t)$ at $t(>\tau)$ is inversely proportional to the square of the radius $r$, $n_\mathrm{e}(r,t)\propto r^{-2}$. 
We obtain the density profile of the ejecta with the mass of $M_\mathrm{ej}$ as
\begin{equation}
n_\mathrm{e}(r,t)=\left\{
\begin{array}{ccl}
\frac{M_\mathrm{ej}}{4\pi \mu m_\mathrm{H}v_\mathrm{ej}\tau r^2}&\mathrm{for}&v_\mathrm{ej}(t-\tau)<r<v_\mathrm{ej}t,\\
0&&\mathrm{otherwise},
\end{array}\right.
\label{dens}
\end{equation}
where $\mu(=1.4)$ and $m_\mathrm{H}(=1.66\times 10^{-24}\ \mathrm{g})$ are the mean molecular weight and the atomic mass unit. 
The velocity of the ejecta is assumed to be $v_\mathrm{ej}=4000$ km s$^{-1}$ from the optical spectroscopy of V2491 Cygni \citep{t08,t08b}. 
The mass of $^{22}$Na produced in ONeMg novae is calculated by \cite{w99} for wide ranges of the mass of the white dwarf and the envelope. 
To produce a sufficient amount of $^{22}$Na, the mass of the envelope is required to be $10^{-3}\ M_\odot$.  
Thus we assume the ejecta mass to be $M_\mathrm{ej}=10^{-3}\ M_\odot$. 
The other parameter characterizing the density structure of the ejecta is the duration $\tau$, which determines the inner radius $R_\mathrm{in}$ of the ejecta, $R_\mathrm{in}=v_\mathrm{ej}(t-\tau)$. 
In the radiative transfer calculation, the inner radius $R_\mathrm{in}$ at $t=t_0(=9\ \mathrm{days})$ is a free parameter to be selected to reproduce the observed X-ray spectrum. 
Using these values, we calculate the optical depth for each photon at $t$. 
The spectrum of the X-ray emission from V2491 Cygni exhibits thermal emission from optically thin plasma with the temperature of $k_\mathrm{B}T=3$ keV \citep{t09}. 
It seems that only the shocked gas located near the boundary between the ejecta and the interstellar medium emits the thermal emission. 
Therefore, the temperature in the rest of the ejecta is much smaller than $k_\mathrm{B}T=3$ keV. 
If the gas in the entire ejecta emitted X-ray photons, the X-ray flux would be much larger than the observed flux.

\subsection{$\gamma$-ray line emission}
Next, the spectrum of seed photons must be specified. 
Following previous works \citep{l92,g98}, we consider the radioactive decay of $^{22}$Na as the dominant source of the $\gamma$-ray line emission. 
The isotope $^{22}$Na decays to a stable isotope $^{22}$Ne by $\beta^+$-decay and electron capture with the half-life of $\tau_{1/2}=2.6$ yr. 
A positron and a $\gamma$-ray photon with the energy of 1.27 MeV are produced by this process. 
The emission from the positron should be treated carefully, because it can be thermalized and form positronium in the ejecta. 
At temperatures below $10^6$ K, the positron prefers the formation of a positronium rather than the direct annihilation to two $\gamma$-ray photons \citep{c76}. 
Thus, the previous works \citep{l87,g98} assumed that 90 $\%$ of positrons produced by $\beta^+$-decay form positroniums. 
In these studies, the positrons were assumed to form positronium after they penetrate a medium with the column mass density of $\sim 0.1$ g/cm$^2$, which is needed for the positrons to thermalize to the energies of $\sim 100$ eV via Coulomb scattering by ions. 
Because the cross section of Coulomb scattering of a positron with the energy of $100$ eV to be $\sim 10^{-18}$ cm$^2$, the number of scattering for the thermalization is on the order of $10^{10}$, which is much larger than the expected number of scattering of a $\gamma$-ray photon (the optical depth of the ejecta on day 9 is $48$). 
Thus, it takes $2\times 10^3$ times longer time for a positron to form positronium than for a photon to escape from the ejecta. 
This means that the flux of the $\gamma$-ray emission is dominated by the direct annihilation. 
Therefore, we can assume that the other 10\% of positrons produced by $\beta^+$-decay are directly annihilated and each of them emits two 511 keV photons. 
The photon fluxes $F_\mathrm{1.27MeV}$ of 1.27 MeV photons and $F_\mathrm{511keV}$ of 511 keV photons from the decay of $^{22}$Na are expressed as,
\begin{eqnarray}
F_\mathrm{1.27MeV}&=&5F_\mathrm{511keV}\nonumber\\
&=&8.6\times 10^{-4}\left[\frac{X(^{22}\mathrm{Na})M_\mathrm{ej}}{3\times10^{-5}\ M_\odot}\right]
\left(\frac{d}{10.5\ \mathrm{kpc}}\right)^{-2}\\
&&\times\exp[-t/\tau(^{22}\mathrm{Na})]
\ \ \mathrm{photons\ cm}^{-2}\ \mathrm{s}^{-1},\nonumber
\label{Fg}
\end{eqnarray}
where $d$ is the distance to the nova, $t$ is the time measured from the onset of the outburst, and $\tau(^{22}\mathrm{Na})=\tau_{1/2}/\mathrm{ln}2$($=3.75$ yr) is the e-folding time of the $\beta^+$-decay. 
The mass fraction of $^{22}$Na synthesized in the ejecta is $X(^{22}\mathrm{Na})$. 
Here $^{22}$Na is assumed to be synthesized immediately after the onset of the outburst. 

In the radiative transfer calculation, we inject $\gamma$-ray photons at a constant rate, because the photon fluxes (\ref{Fg}) are nearly constant for the time scale considered here. 
For 1/6 of the seed photons, the initial energies are set to be $511$ keV. 
The others have the initial energies of $1.27$ MeV. 
The initial position of each photon is determined by a random number so that the spatial distribution of the seed photons follows the density distribution (\ref{dens}) in the ejecta. 
The $\gamma$-ray line fluxes are assumed to be $F_\mathrm{1.27MeV}=8.6\times 10^{-4}$ photons cm$^{-2}$ s$^{-1}$ and $F_\mathrm{511keV}=1.7\times 10^{-4}$ photons cm$^{-2}$ s$^{-1}$. 

\begin{figure}
\begin{center}
\includegraphics[scale=0.5]{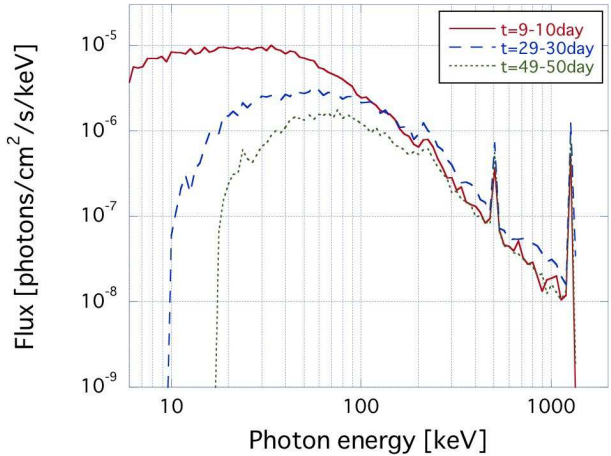}
\caption{Resultant energy spectra of photons escaping from the ejecta.
The seed photons are injected from $t=9$ day to $t=10$ day (solid line) , from $t=29$ day to $t=30$ day (dashed line), and from $t=49$ day to $t=50$ day (dotted line). 
The free parameters characterizing the ejecta are $v_\mathrm{ej}=4000$ km s$^{-1}$, $M_\mathrm{ej}=10^{-3}\ M_\odot$, and $R_\mathrm{in}=3\times 10^{12}$ cm. 
Photoelectric absorption is neglected ($Z=0$). }
\label{f1}
\end{center}
\end{figure}

\subsection{Radiative processes}
Compton scattering and photoelectric absorption are taken into account as follows. 
For Compton scattering, we simply use the Thomson cross section $\sigma_\mathrm{T}$($=6.65\times 10^{-25}$ cm$^2$). 
Using the analytical formula given by \cite{v96}, the photoionization cross section for an incident photon with the energy $E$ is expressed as
\begin{equation}
\sigma_\mathrm{ph}(E)=\left\{
\begin{array}{ccl}
\sigma_0\frac{(x-1)^2x^{0.5P-5.5}}{(1+\sqrt{x/y_\mathrm{a}})^{-P}}&\mathrm{for}&E>E_\mathrm{th},\\
0&\mathrm{otherwise,}&
\end{array}\right.
\label{sigma}
\end{equation}
where $x$ is the energy of the incident photon normalized by an energy $E_0$,
\begin{equation}
x=\frac{E}{E_0}.
\end{equation}
It is difficult to specify the dominant opacity source due to unknown ionization states in the ejecta. 
Since the spectrum of the X-ray emission on day 9 \citep{t09} exhibits the K$\alpha$ line from \ion{Fe}{25}, we use the values of parameters in Equation (\ref{sigma}) corresponding to photoelectric absorption by \ion{Fe}{25}. 
That is, $E_0=1.057$ keV, $\sigma_0=1.195\times 10^{-17}$ cm$^2$, $y_\mathrm{a}=57.69$, and $P=1.718$. 
The ionization potential is given by $E_\mathrm{th}=8.829$ keV. 

We introduce a parameter $Z$ as the number ratio of \ion{Fe}{25} ions to electrons in the ejecta. 
Assuming a matter with the solar abundance and that all iron is in the form of \ion{Fe}{25},  the value of $Z$ becomes $\sim 2\times10^{-5}$ . 
Using this parameter set, we can express the total cross section $\sigma$ as,
\begin{equation}
\sigma=\sigma_T+Z\sigma_\mathrm{ph}(E),
\end{equation}
and the probability $P_\mathrm{ph}$ of photoelectric absorption as 
\begin{equation}
P_\mathrm{ph}=\frac{Z\sigma_\mathrm{ph}(E)}{\sigma_\mathrm{T}+Z\sigma_\mathrm{ph}(E)},
\end{equation}
In the 
calculation, we evaluate the probability of absorption or scattering of each photon using the mean free path of the photon with respect to the total cross section. 
When a photon interacts with the matter, we generate a random number ranging from $0$ to $1$.  
The photon is scattered by an electron if the number is greater than $P_\mathrm{ph}$. 
Otherwise the photon is absorbed and we stop tracing this photon. 

 \begin{figure}
\begin{center}
\includegraphics[scale=0.5]{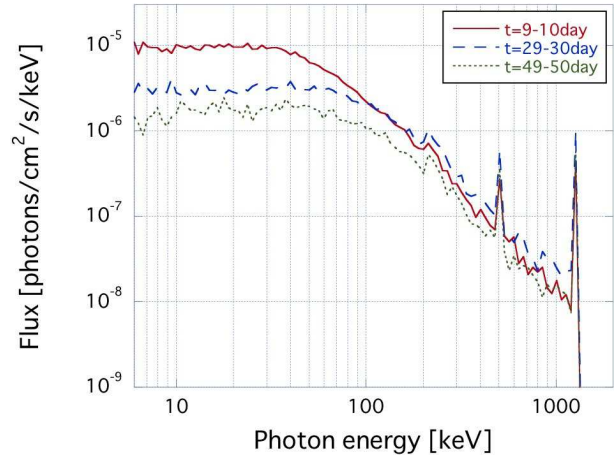}
\caption{Same as Figure 1, but for $R_\mathrm{in}=3\times 10^{11}$ cm.}
\label{f2}
\end{center}
\end{figure}

\section{RESULTS\label{results}}
In this section, we show results of the radiative transfer calculation. 
The total number of photons used in the calculation is 30000.

\subsection{Spectrum}
Figure \ref{f1} shows the resultant spectra of photons escaping from the ejecta in which photoelectric absorption is neglected ($Z=0$). 
The photons are injected from $t=9$ day to $t=10$ day (solid line, referred to as period I below), from $t=29$ day to $t=30$ day (dashed line, referred to as period II), and from $t=49$ day to $t=50$ day (dotted line). 
The parameters characterizing the evolution of the ejecta are $v_\mathrm{ej}=4000$ km s$^{-1}$, $M_\mathrm{ej}=10^{-3}\ M_\odot$, and $R_\mathrm{in}=3\times 10^{12}$ cm. 
With this parameter set, the optical depth of the ejecta is $\tau=48$ on day 9
and $\tau=0.065$ on day 29. 
At first, one can easily recognize two spikes at $511$ keV and $1.27$ MeV in the spectrum.
They are photons without being scattered in the ejecta. 
Except for the two spikes, the spectrum is continuous, which is a result of comptonization of the $\gamma$-ray photons. 
Especially, there is an extremely flat part ($10$ keV$<E<50$ keV) in the spectrum of photons accumulated for period I. 
The spectral analysis of the X-ray emission of V2491 Cygni shows that the spectrum in the energy range of $10$ keV $<E<70$ keV is well fitted by a power law with the photon index of $0.1\pm 0.2$ \citep{t09}. 
This feature is well reproduced by our model. 
The analytical investigation of comptonization of photons escaping from a spherical plasma cloud by \cite{st80} found similar flat spectra. 
The flat spectrum in Figure \ref{f1} seems to be formed in the same manner. 
In addition, the flux in the flat part is consistent with the observed value. 
Thus, our model successfully reproduces the observed spectrum of the hard X-ray emission from V2491 Cygni on day 9. 

On the other hand, the observation on day 29 detected no hard X-ray emission. 
The spectrum of period II shows a significant decrease of the flux in the energy range of $10$ keV $<E<70$ keV. 
This is a result of the expansion of the ejecta. 
The decreasing optical depth allows a large fraction of $\gamma$-ray photons to escape from the ejecta without being scattered. 
Thus, our model can explain the absence of the hard X-ray emission on day 29. 

This fast change of the X-ray emission is in contrast to a similar model by \cite{l92} in which the timescale of the X-ray light curve is of the order of 100 days or more. 
This difference comes from the assumed expansion velocities of the ejecta in the two models. 
\cite{l92} assumed $v_\mathrm{ej}= 10$ km s$^{-1}$ or 6 km s$^{-1}$ while we assume a much higher velocity $v_\mathrm{ej}= 4000$ km s$^{-1}$ as suggested from optical observations for this particular nova.

\begin{figure}
\begin{center}
\includegraphics[scale=0.5]{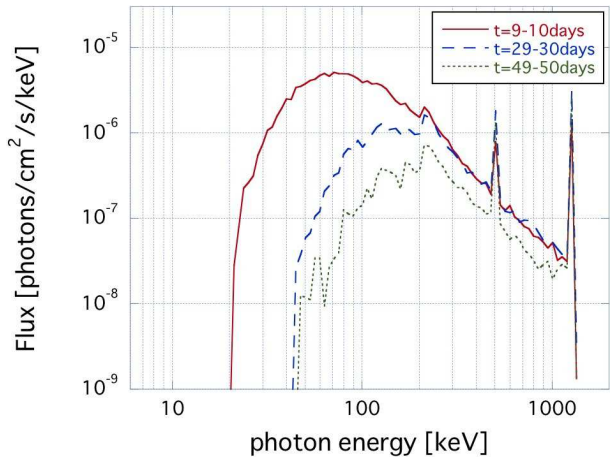}
\caption{Same as Figure 1, but for $R_\mathrm{in}=3\times 10^{13}$ cm.}
\label{f4}
\end{center}
\end{figure}

\subsection{Effects of the inner radius}
Figure \ref{f2} shows the resultant spectra of the model with $R_\mathrm{in}=3\times 10^{11}$ cm. 
As well as the model with $R_\mathrm{in}=3\times 10^{12}$ cm, the flat spectrum is obtained in period I. 
However, the spectrum for period II remains flat. 
We can attribute this feature to effects of the small $R_\mathrm{in}$. 
For the model with a small $R_\mathrm{in}$, the fraction of photons travelling in the deep interior of the ejecta is large compared to that with a large $R_\mathrm{in}$. 
Since low-energy photons are produced by repeated Compton scattering in the deep interior, the fraction of low-energy photons remains large even on several tens days after the outburst. 
Although the value of the flux in the flat part of the spectrum for period II is smaller than that of period I, it may contradict the non-detection of the hard X-ray emission on day 29. 
On the other hand, Figure \ref{f4} shows the resultant spectra of the model with $R_\mathrm{in}=3\times 10^{13}$ cm. 
In this case, the spectrum of period I fails to reproduce the flat spectrum. 
This is because the density of the ejecta becomes already too low to remain opaque to the $\gamma$-ray line in the first 10 days. 
As a consequence, the model with $R_\mathrm{in}=3\times 10^{12}$ cm is preferred to explain the observational features of the X-ray emission from V2491 Cygni.  
Therefore, the mass ejection from the stellar surface is terminated by day 10.

\subsection{Effects of photoelectric absorption}
\begin{figure}
\begin{center}
\includegraphics[scale=0.5]{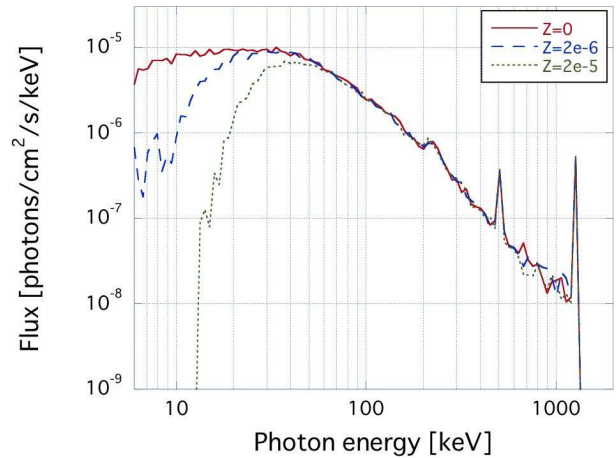}
\caption{Energy spectra of photons injected from $t=9$ day to $t=10$ day with $Z=0$ (solid line), $2\times 10^{-6}$ (dashed line), and $2\times 10^{-5}$ (dotted line).  }
\label{f3}
\end{center}
\end{figure}

Figure \ref{f3} shows the resultant spectra for various values of the ratio $Z$ of the number of  \ion{Fe}{25} ions to that of electrons in the ejecta. 
The seed photons are injected from $t=9$ day to $t=10$ day. 
Each line represents the model with $Z=0$ (solid line), $Z=2\times 10^{-6}$ (dashed line), and $Z=2\times 10^{-5}$ (dotted line). 
For the solar abundance, the number ratio of Fe to H is $2\times 10^{-5}$. 

One can recognize the tendency that the photon flux at several tens keV decreases with the increasing ratio $Z$. 
Even for 1/10 of the solar value ($Z=2\times 10^{-6}$), the flat part of the spectrum to be seen in the model with $Z=0$ disappears. 
Therefore, the model with a small amount of heavy elements or the almost fully ionized envelope is appropriate to account for the flat spectrum of V2491 Cygni. 

\section{DETECTABILITY OF GAMMA-RAY LINE}
The ejecta become transparent to the $\gamma$-ray line emission within several tens days. 
Therefore, the $\gamma$-ray line emission may be detected by $\gamma$-ray observations. 
In this section, we discuss the detectability of the $\gamma$-ray line emission from V2491 Cygni. 
Results of the calculation 
show that the initial photon fluxes of the $\gamma$-ray line emissions are required to be $F_\mathrm{1.27MeV}=8.6\times10^{-4}$ photons cm$^{-2}$ s$^{-1}$ and $F_\mathrm{511keV}=1.7\times 10^{-4}$ photons cm$^{-2}$ s$^{-1}$ to explain the hard X-ray emission from V2491 Cygni. 
A few years after the discovery of V2491 Cygni, the photon fluxes must decrease by a factor $\sim 2$ because of the half-life of $\tau_{1/2}=2.6$ yr. 
Thus, the present value of the photon fluxes are estimated as $F_\mathrm{1.27MeV}\sim 4\times 10^{-4}$ photons cm$^{-2}$ s$^{-1}$ and $F_\mathrm{511keV}\sim 9\times 10^{-5}$ photons cm$^{-2}$\ s$^{-1}$. 
To detect these $\gamma$-ray lines with sufficient S/N ratios ($\sim 5$), about 100 ksec observations are required both for the {\it INTEGRAL} spectrometer SPI and the {\it Fermi} Gamma-ray Burst Monitor. 

\section{CONCLUSIONS\label{conclusions}}
In this Letter, we consider the possibility that Compton degradation of the $\gamma$-ray line emission from the radioactive isotope $^{22}$Na can explain the observed hard X-ray emission from the classical nova V2491 Cygni. 
We adopt a simple wind model as the ejecta model and calculate radiative transfer of the $\gamma$-ray line emission in the ejecta. 
As a result, we succeed in reproducing the spectrum of the hard X-ray emission on the 9th day after the discovery. 
At the same time, our model can explain the absence of the hard X-ray emission on day 29. 
This is because the optical depth of the ejecta decreases to 0.065 in these 20 days. 
The amount of $^{22}$Na synthesized in the ejecta is required to be $3\times 10^{-5}\ M_\odot$ to account for the flux of the hard X-ray emission.  
We also estimate the present value of the photon fluxes of the 1.27 MeV and 511 keV line emissions and find that these line emissions can be detected by the GBM on {\it Fermi} and the SPI on INTEGRAL with a reasonable observing time. 

Finally, we mention an inadequacy of our model. 
Previous observations and theoretical investigations of the outbursts on massive white dwarfs imply the ejecta mass of $10^{-5}\ M_\odot$, which is much smaller than that of our model. 
In this case, the radioactive decay of $^{22}$Na must not be the origin of the hard X-ray emission from V2491 Cygni, because the amount of $^{22}$Na is much smaller than that required to reproduce the emission. 
However, the X-ray detection from the pre- and post-outburst images of V2491 Cygni may imply that it was an unusual nova. 
Therefore, observations of the $\gamma$-ray line emission by the {\it INTEGRAL} and/or the {\it Fermi} must provide us crucial information on the hard X-ray emission.

\acknowledgments
This work has been partly supported by  Grant-in-Aid for JSPS Fellows (21$\cdot$1726) and Grants in Aid for Scientific Research (21018004) of the Ministry of Education, Science, Culture, and Sports in Japan.

\end{document}